# Mass distribution in 36.2 MeV alpha induced fission of $^{232}$Th

D. Banerjee[1], T. N. Nag[2], R. Tripathi[2], Sk Wasim Raja[1],
S. Sodaye[2], P. K. Pujari[2]

[1]Radiochemistry Division (BARC), Variable Energy Cyclotron Centre, 1/AF, Bidhan Nagar, Kolkata-700064

[2]Radiochemistry Division, Bhabha Atomic Research Centre, Mumbai-400085

[3]A. Chakrabarti, [3]M. Bhattacharjee, [3]L. K. Doddi, [3]V. Naik

[3]Radioactive Ion Beam Group, Variable Energy Cyclotron Centre and HBNI-Kolkata, 1/AF, Bidhan Nagar, Kolkata-700064

**Abstract**

Mass distribution of fission products has been determined in ($\alpha$+$^{232}$Th) reaction at $E_{lab}$=36.2 MeV using $\alpha$ particles from the cyclotron at the Variable Energy Cyclotron Centre (VECC), Kolkata. Yields of 69 fission products having half-lives in the range of about ~1 min to several days have been measured using gamma ray spectrometry of fission products. The mass distribution obtained on the basis of the yields shows a clear triple humped structure showing the contribution from both asymmetric and symmetric modes of fission. Comparison of the experimental mass distribution with the prediction based on the GEF code, which takes into account multi-chance fission, shows that the simulation based on GEF reasonably reproduces asymmetric component arising from standard I mode of fission, but underestimate the contribution from Standard II fission mode. Also, the peak like structure in the symmetric region could not be reproduced.

Corresponding author email: dbanerjee@vecc.gov.in

## I. Introduction

Historically, nuclear fission has been described using a macroscopic approach where the potential energy of a liquid drop is traced as a function of deformation of the nucleus undergoing fission [1-3]. However, the observed asymmetric mass distribution in fission of actinides, especially in thermal neutron induced fission of $^{235}$U at low excitation energies could only be explained by incorporating shell effects in the macroscopic liquid drop approach [4]; especially the spherical shell corresponding to N=82 and deformed shell corresponding to N=88 appear to play a major role in the asymmetric split. The shell effect is expected to play a role only at low excitations and one expects a gradual washing out of the shell effect with increasing excitation [5-7], resulting in a gradual shift from asymmetric to a completely symmetric split. It has been found that in 14 MeV neutron induced fission of $^{235}$U, the asymmetric peak to symmetric valley ratio in the mass distribution decreases to a value of 6 from 600 observed in thermal neutron induced fission due to the decrease in strength of shell effects at higher excitation energy. However, excitation energy region, where shell effects are completely washed out, is still not fully established experimentally.

The Compound nucleus, $^{236}$U, can be produced in (n+$^{235}$U) as well as in (α+$^{232}$Th) reaction. It is easy to get α-particles at different energies from different types of accelerators and a number of studies on fission product mass distribution at higher excitations of the $^{236}$U compound nucleus have been carried out by different groups using the α+ $^{232}$Th reaction [8-12]. In two of the earlier studies the mass distributions in $^{232}$Th $(\alpha, f)$ reaction were found to have three peaks/humps [8,9]. Roginski et al, [8] observed triple hump mass distribution, corresponding to asymmetric and symmetric splitting, for incident α- particle energies of 33 and 39 MeV, whereas for incident energy of 22.9 MeV, double hump asymmetric mass distribution was observed. Chakrabarti et al. [9] observed that the mass distribution has three peaked (or triple humped) structure over a wide range of $\alpha$ particle energy from 28.5 to 71.4 MeV. However, there are studies that observed either double humped mass distribution or completely symmetric mass distribution. For example, Guin et.al [10], reported double humped asymmetric mass distribution for the same system for incident $\alpha$ particle energy 28.5 MeV; Chaudhuri, et.al [11], have reported complete washing out of shell effects beyond 40 MeV leading to a symmetric mass distribution. Also, a recent study carried on a number of actinide fissioning systems close to $^{236}$U (e.g. $^{237-240}$U, $^{239-242}$Np, etc.), over an excitation energy range of 10 to 60 MeV shows asymmetric double humped mass distribution up to 60 MeV of excitation [12] but no peak in the symmetric region.

Three peaked mass distributions were, however, consistently observed in the fission of $^{228}$Ra at low excitation [13], which signifies the co-existence of multi-mode fission of the same fissioning nucleus. Only rather recently, a suitable explanation for this observation has been put forward by Moller et al, by describing the shape evolution of the fissioning nucleus on a five dimensional potential energy surface [14]. Apart from elongation, neck diameter and mass asymmetry, the authors considered two more shape parameters; the deformations of the left and right nascent fragments. With this prescription they were able to explain the observed two mode fission in $^{228}$Ra. But their calculations do not predict bi-modal fission for Uranium nuclei.

As the excitation energy increases, the probability of multi-chance fission [15] increases. In multi-chance fission (MCF), fission occurs after neutron evaporation. Thus first chance fission means fission of the original compound nucleus ($^{236}$U for α + $^{232}$Th system), second chance fission for the same projectile-target system means fission of $^{235}$U after one neutron evaporates out of $^{236}$U; third chance means fission of $^{234}$U after consecutive evaporations of two neutrons, and so on till the excitation energy of the compound nucleus after successive emission of neutrons falls below the fission barrier. Since neutron evaporation takes away on the average excitation energy that exceeds the binding energy of the neutron by about 1-2 MeV (typically close to ~7.5 MeV in this case), the higher chance fissions proceed increasingly from lower excitations of the compound systems. Depending upon the initial excitation, after evaporation of a few neutrons the excitation of the compound system would come down to a level where shell effects play an important role leading to an asymmetric split. Multi-chance fission can thus explain the survival of asymmetric mass distribution to high excitation energies with excitations reaching 60 MeV and a little more, as observed in earlier experimental studies [9, 12].

In view of these new theoretical developments and experimental findings, the accurate determination of mass distribution at comparatively higher excitations in the fission of actinides has gained renewed interest. Also, the mass distribution data at higher excitations is important for ADS development. A fresh effort to determine accurately the mass distribution of actinide fissioning systems at different excitations is therefore necessary and it would be interesting to verify and explain, if confirmed, the previously observed triple peaked mass distribution in $\alpha$ induced fission of $^{232}$Th.

The present study deals with the determination of the mass distribution in $\alpha$ induced fission of $^{232}$Th for the incident $\alpha$ particle energy of 36.2 MeV, which corresponds to 31 MeV excitation of the compound nucleus $^{236}$U. The He-jet recoil transport system at VECC [16] has been used for the transport of the fission products from the target chamber to a low background detection site, about 12 m away. The fission yields (FYs) of 69 fission products (FPs) were measured, which allowed a reliable determination of the mass distribution. The experimental mass distribution is compared with the calculation of GEF code [17].

## 2. Experimental details

The experiment was carried out at Variable Energy Cyclotron Centre (VECC), Kolkata. A self-supporting target of $^{232}$Th (thickness: 8 mg/cm$^2$) was irradiated with 40 MeV $\alpha$ beam from the $K$=130 AVF Cyclotron. The projectiles entered into the irradiation chamber through a 25 micron thick Havar window and then passed through a 25 µm super pure aluminum foil placed upstream of the target before bombarding the target. The energy loss of the projectile in the Havar window, in the aluminum foil, and in the target was calculated using the software SRIM [18] and resulted in $\alpha$ beam energy of 36.2 MeV after traversing half the thickness of the $^{232}$Th target. The irradiations were performed for three different durations: 2min, 10min and 2h denoted by R-I, R-II and R-III respectively. All through the irradiation, the beam intensity was monitored every 10 s intervals to take care of the fluctuation in the beam intensity during irradiation, if any. The reaction products recoiling out of the target were transported by He-jet system to a low-background counting area and implanted on a graphite catcher foil. The catcher foil was then counted with a pre-calibrated high purity germanium (HPGe) detector for a period of about 30 min for R-I, ~2 h for R-II and for a few days for R-III. Different irradiation times helped in determining the yields of fission products with half-lives spanning over a very wide range. The data acquisition was performed by PC-based, software controlled PCI-bus multichannel analyzer FAST ComTec MCA-3 [19].

A typical γ-ray spectrum of the fission products obtained in R-I is shown in Fig. 1 The figure also shows the assignment of different γ-lines in the spectrum to the various fission products. The assignments were made on the basis of γ-ray energies, ensuring at the same time that the decay behaviors were consistent with the reported half-lives. The decay data of the FPs used in the present study was taken from the literature [20] and is given in Table 1. The γ-ray spectra were analyzed using the peak area analysis software PHAST [21]. The peak areas under the characteristic γ-rays of different fission products were used to obtain their 'end of irradiation activities'. The end of irradiation activities were used to obtain the yields of the fission products using the procedure as discussed earlier [22,23].

## 3. Results and Discussion:

The yield of a mass chain $Y(A)$ is obtained from experimentally measured independent yield $IN(A,Z)$ or cumulative yield $CY(A,Z)$ of a fission product with mass $A$ and atomic number $Z$, using the following equations:

$$Y(A) = \frac{IN(A,Z)}{FIY(A,Z)} \quad \text{...................(1)}$$

$$Y(A) = \frac{CY(A,Z)}{FCY(A,Z)} \quad \text{...................(2)}$$

Where $FIY(A,Z)$ and $FCY(A,Z)$ are fractional independent and cumulative yields respectively and are calculated using the equations:

$$FCY(A,Z) = \frac{1}{\sqrt{2\pi\sigma_z^2}} \int_{-\infty}^{z+0.5} e^{-(z-z_p)^2/2\sigma_z^2} dz \quad \text{............(3)}$$

$$FIY(A,Z) = \frac{1}{\sqrt{2\pi\sigma_z^2}} \int_{z-0.5}^{z+0.5} e^{-(z-z_p)^2/2\sigma_z^2} dz \quad \text{............(4)}$$

The most probable Z ($Z_p$) for a given mass chain $A$ and the width of the isobaric yield distribution ($\sigma_z$) are two parameters which are required to carry out the charge distribution correction to obtain the mass yield $Y(A)$ from the experimentally measured yields of the respective fission products [24]. In this work, $\sigma_z=0.7$ has been used for the charge distribution correction as was used by Umezawa et al. [25] for the fission of different actinide nuclei including $^{236}$U in the similar excitation energy range. The $Z_P$ value for a mass chain with mass number $A$ was calculated as $A/[(A_{CN} − \nu_T)/Z_{CN}]$, where $A_{CN}$ and $Z_{CN}$ are, respectively, the mass number and proton number of the compound nucleus (CN) and $\nu_T$ is the average number of neutrons emitted per fission calculated using the prescription of Kozulin et al. [26].

Determination of absolute yields (cumulative/independent yields) requires He-jet transport efficiency. However, to avoid the uncertainties regarding the constancy of He-jet transport efficiency in between the different runs, the ratio of total fission cross-section calculated using the code PACE2 [27] and the experimental fission cross-sections (area under the mass distribution curves) has been taken as the normalization factor to determine the present yields as given in Table I. The MD curve obtained after combining the mass yield data from all the three irradiations is shown in Fig. 2. The yields obtained from multiple runs for a particular isotope have been averaged and that average yield has been used for finding the mass yield. The mass distribution was fitted using three Gaussian functions, one each for two asymmetric regions and one for the symmetric region. The centroid of the symmetric fission is at 115.2±0.2, which corresponds to a fissioning system of mass number 230. This implies loss of 6 neutrons from the Compound nucleus ($^{236}$U) which is consistent with the prescription of Kozulin et.al [26]. The asymmetric peaks are centered at 98.4 ±0.1 and 136.7 ±0.1 respectively.

The present study clearly shows that the mass distribution has a three peaked structure with two peaks for asymmetric splitting and one for symmetric splitting. Thus the mass distribution has contribution both from the asymmetric and symmetric fission modes. The asymmetric peaks are much more pronounced indicating that asymmetric fission is the dominant mode. The dominance of the asymmetric fission mode is consistent with some of the earlier studies [8, 9] and one recent study [12] but not with the findings of Chaudhuri et al,[11] who reported nearly flat top mass distribution in the similar excitation energy range with dominant contribution from symmetric fission. However, in the recent study [12], the authors did not observe the symmetric peak but found that fission of actinide systems in the neighbourhood of $^{236}$U remains asymmetric up to 60 MeV of excitation energy, the maximum excitation studied in the experiment.

In order to investigate if the multi-chance fission can explain the observed three peaked mass distribution, the mass distribution was calculated using the GEF code [17]. The calculated mass distribution was normalized to experimental mass distribution using the mass yield data in 80-150 mass regions. The GEF calculation includes mass distribution arising from the asymmetric fission modes standard I (corresponding to N=82) and standard II (corresponding to N=88), and symmetric fission mode. It can be seen from figure 2 that the calculated mass distribution, although qualitatively is in fair agreement with the experimental mass distribution, does not reproduce the symmetric peak satisfactorily. Also, the contribution from the asymmetric mode corresponding to the N=88 shell is underestimated as compared to the experimental data. Nevertheless, it is evident both from experimental data and theoretical calculations, that the shell effects play a significant role in the fission of $^{236}$U at excitation energy of 31 MeV. Apparently, it is the contribution from multi-chance fission that results in pronounced shell effect. The GEF code estimates the contribution from first, second, third and fourth chance fission to be 22%, 30%, 38% and 9.6% respectively. The corresponding excitation energies of the fission nuclei were 31 MeV, 23 MeV, 16 MeV and 8.5 MeV respectively. Nevertheless, it is evident from both experimental data and theoretical calculations, that the shell effects are strongly pronounced for the fission of $^{236}$U at excitation energy of 31 MeV. This may possibly be due to the contribution from multi-chance fission. Except for the first chance fission the other excitation energies are the approximate average values. Thus, about ~78% of fission is occurring with excitation energy of about ~23 MeV or less which can result in significant shell effects in the fission process.

## 4. Conclusions

Measurement of mass distribution has been carried out in α+$^{232}$Th reaction at $E_{lab}$=36.2 MeV. The use of He-gas jet transport and different irradiation times allowed determination of yields of a large number (69) of fission products having half-lives spanning over a wide range starting from ~1 min to several days. This has allowed a comprehensive determination of the mass distribution. The mass distribution has been observed to be clearly triple humped with dominant contribution from asymmetric fission. This observation is consistent with some but not all of the earlier measurements carried out on the same system by various experimental groups. It is found that the calculations based on the GEF code, which takes into account the multi-chance fission (MCF), does not show a clear three peak structure, although it can explain the survival of the asymmetric fission.

**Acknowledgement:** We would like to thank Shri P. S. Chakraborty and operations team of *K*130 Cyclotron at VECC for their contribution in beam-delivery for the experiments.

**Captions for the table**

| | |
|---|---|
| Table 1 | Decay data [20] and Yields of fission products identified in the present study for $\alpha+{}^{232}$Th reaction at $\alpha$ particle energy of 36.2 MeV (See text for determination of absolute yields). The symbols 'C' and 'I' represents cumulative and independent yields respectively. |
| Captions for the figures | |
| Fig. 1. | A typical γ-ray spectrum of the fission products obtained after 2 min irradiation followed by 30 s counting. Gamma-rays due to the different fission products are labeled. |
| Fig. 2. | Mass distribution for α + ${}^{232}$Th reaction at $E_\alpha$ = 36.2 MeV. Open circles are experimental yields (see text for the determination of absolute yields). Black solid line shows the overall fit using sum of three Gaussian function including contribution from both symmetric and asymmetric fission. Green line is the mass distribution calculated using GEF code [17]. |

**Table I:** Decay data [20] and Yields of fission products identified in the present study for $\alpha+^{232}$Th reaction at $\alpha$ particle energy of 36.2 MeV (See text for determination of absolute yields). The symbols 'C' and 'I' represents cumulative and independent yields respectively.

| Sl. No. | Nuclide | Half-life | $E_\gamma$ (keV) | Yield (mb) |
|---|---|---|---|---|
| 1 | $^{84}$Se | 3.26m | 408.2 | 9.8±1.9 (C) |
| 2 | $^{89}$Rb | 15.4m | 657.7 | 10.4±1.0 (C) |
| 3 | $^{90m}$Rb | 4.3m | 831.7 | 13.9±1.4 (C) |
| 4 | $^{90g}$Rb | 2.63m | 831.7 | 41±6 (C) |
| 5 | $^{91}$Sr | 9.63h | 749.8 | 40.1±1.8 (C) |
| 6 | $^{92}$Sr | 2.71h | 1383.9 | 37.1±1.5 (C) |
| 7 | $^{93}$Y | 10.18h | 266.9 | 41±4 (C) |
| 8 | $^{94}$Sr | 1.255m | 1427.7 | 28.7±2.6 (C) |
| 9 | $^{94}$Y | 18.6m | 918.8 | 30.5±1.6 (C) |
| 10 | $^{95}$Y | 10.3m | 954 | 42.2±2.9 (C) |
| 11 | $^{95}$Zr | 64.02d | 756.7 | 46±6 (C) |
| 12 | $^{97}$Zr | 16.744h | 657.9 | 49.1±1.1 (C) |
| 13 | $^{99}$Mo | 65.94h | 140.5 | 50.6±1.2 (C) |
| 14 | $^{101}$Mo | 14.61m | 590.9 | 55.1±3.3 (C) |
| 15 | $^{102}$Mo | 11.3m | 148.2 | 59±7 (C) |
| 16 | $^{103}$Ru | 39.25d | 497.1 | 39.5±1.5 (C) |
| 17 | $^{105}$Ru | 4.44h | 724.2 | 35.2±0.9 (C) |
| 18 | $^{105}$Rh | 35.36h | 318.9 | 37.4±1.2 (C) |
| 19 | $^{105}$Tc | 7.6m | 143.26 | 27±6 (C) |
| 20 | $^{106}$Tc | 0.59m | 270.1 | 20±6 (C) |
| 21 | $^{107}$Ru | 3.75m | 194.3 | 23±4 (C) |
| 22 | $^{107}$Rh | 21.7m | 302.8 | 12.8±1.0 (C) |
| 23 | $^{109}$Rh | 1.35m | 326.9 | 27±4 (C) |
| 25 | $^{111}$Ag | 7.45d | 342.1 | 30.4±3.4 (C) |
| 26 | $^{112}$Pd | 21.05h | 617.4 | 34.8±1.1 (C) |
| 27 | $^{113g}$Ag | 5.37h | 298.6 | 32.2±1.4 (C) |
| 28 | $^{114}$Pd | 2.42m | 126.7 | 31±11 (C) |
| 29 | $^{115}$Ag | 20m | 229 | 30±3 (C) |
| 30 | $^{115g}$Cd | 2.23d | 336.2 | 29.2±1.0 (C) |
| 31 | $^{117m}$Cd | 3.36h | 552.9 | 14.7±0.8 (C) |
| 32 | $^{117g}$Cd | 2.49h | 273.3 | 14.2±0.8 (C) |
| 33 | $^{119m}$Cd | 2.2m | 1025 | 19.8±1.9 (C) |
| 34 | $^{123}$Sn | 40.06m | 160.32 | 32.1±2.5 (C) |
| 35 | $^{125}$Sn | 9.52m | 331.9 | 12.6±1.5 (C) |
| 36 | $^{126g}$Sb | 12.46d | 666.3 | 7.08±0.53 (C) |
| 37 | $^{127}$Sn | 4.13m | 490.6 | 3.07±0.47 (C) |
| 38 | $^{127}$Sb | 3.85d | 685.7 | 31.2±1.0 (C) |
| 39 | $^{128}$Sn | 59.07m | 482.3 | 14.7±2.4 (C) |
| 40 | $^{128}$Sb | 9.01h | 754 | 18.9±1.0 (C) |
| 41 | $^{129m}$Sb | 17.7m | 759.8 | 9.9±0.8 (I) |
| 42 | $^{129g}$Sb | 4.36h | 914.96 | 22.0±2.0 (C) |
| 43 | $^{130g}$Sb | 39.5m | 793.4 | 17.6±2.0 (C) |
| 44 | $^{130m}$Sb | 6.3m | 793.4 | 9.0±0.3 (C) |
| 45 | $^{130m}$I | 8.84m | 536.1 | 16.1±4.6 (I) |
| 46 | $^{131}$I | 8.02d | 364.5 | 44.1±1.3 (C) |
| 47 | $^{133g}$Te | 12.5m | 312.1 | 13.1±1.5 (C) |
| 50 | $^{133g}$I | 20.8h | 529.9 | 36.4±1.1 (C) |
| 51 | $^{133m}$Te | 55.4m | 334.3 | 24.25±1.66 (C) |
| 52 | $^{134g}$I | 52.5m | 847 | 25.3±1.3 (C) |
| 53 | $^{134m}$I | 3.52m | 272.1 | 12.2±1.6 (I) |
| 54 | $^{138m}$Cs | 2.91m | 463 | 44±4 (I) |
| 55 | $^{139}$Ba | 83.06m | 165.8 | 32.2±1.7 (C) |
| 56 | $^{140}$Ba | 12.75d | 537.3 | 42.0±2.5 (C) |
| 57 | $^{141}$Ce | 32.5d | 145.4 | 44.4±1.5 (C) |
| 58 | $^{141}$Ba | 18.27m | 190.3 | 35.14±1.92 (C) |
| 59 | $^{142}$Ba | 10.6m | 255.3 | 33.8±1.9 (C) |
| 60 | $^{142}$La | 91.1m | 641.2 | 35.2±1.9 (C) |
| 61 | $^{145}$Ce | 3.01m | 724.33 | 20.1±3.5 (C) |
| 62 | $^{146}$Ce | 13.49m | 316.7 | 22.1±1.2 (C) |
| 63 | $^{147}$Pr | 13.4m | 641.4 | 40.5±4.5 (C) |
| 64 | $^{148}$Pr | 2.01m | 697.5 | 32±7 (I) |
| 65 | $^{149}$Pr | 2.26m | 138.5 | 17.7±1.65 (C) |
| 66 | $^{150}$Pm | 2.68h | 333.9 | 1.44±0.22 (I) |
| 67 | $^{151}$Nd | 12.44m | 116.8 | 3.92±0.46 (C) |
| 68 | $^{152}$Nd | 11.4m | 250.2 | 4.21±0.29 (C) |

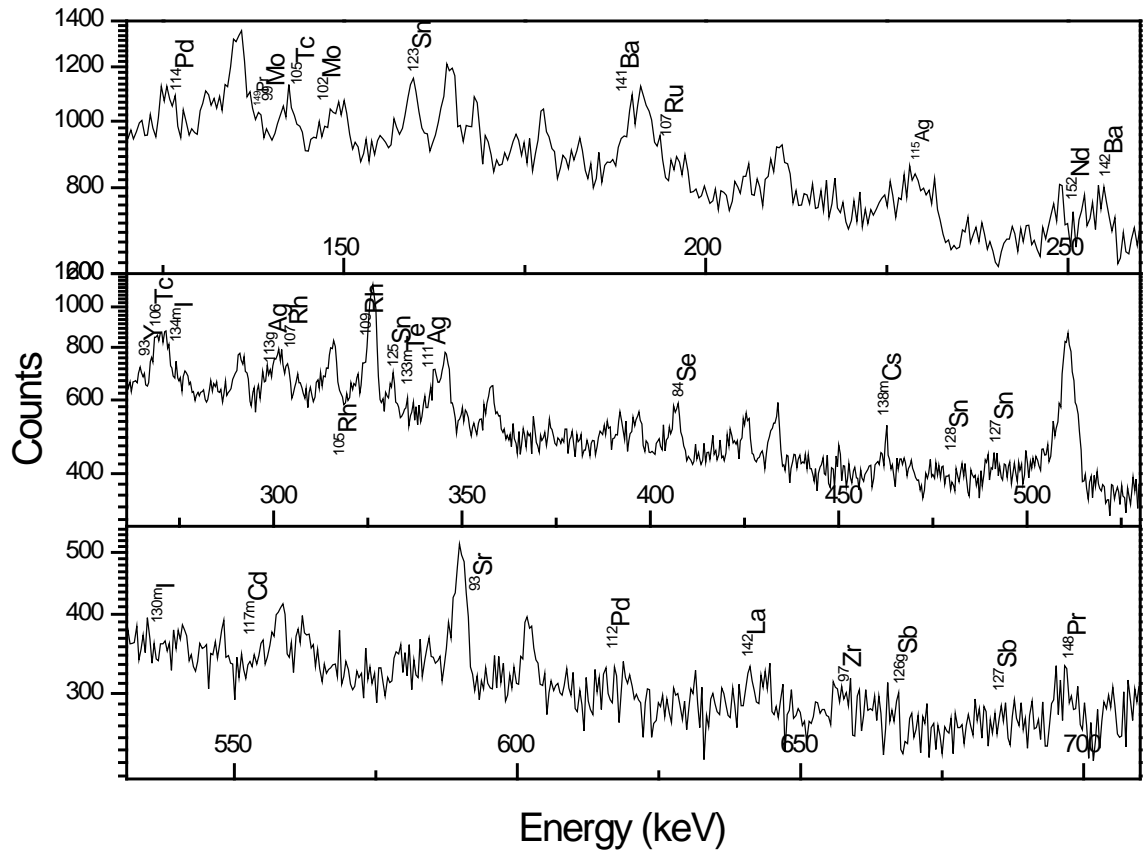

Fig. 1. A typical γ-ray spectrum of the fission products obtained after 2 min irradiation followed by 30 s counting. Gamma-rays due to the different fission products are labeled.

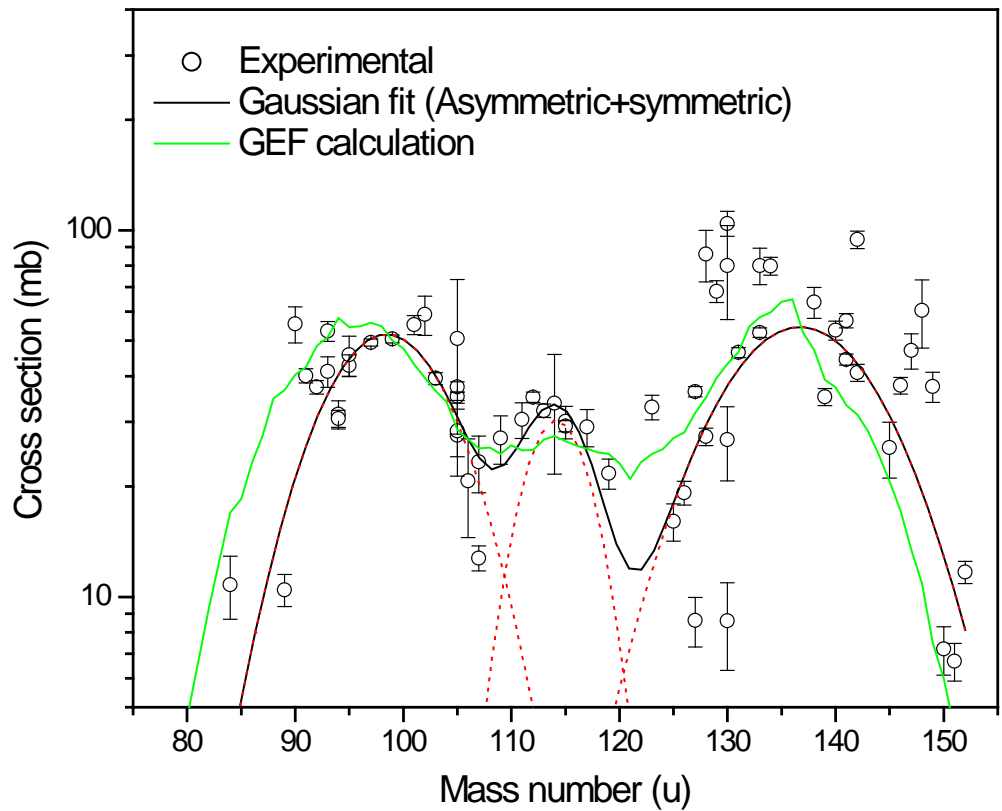

Fig. 2. Mass distribution for α + $^{232}$Th reaction at $E_\alpha$ = 36.2 MeV. Open circles are experimental yields (see text for the determination of absolute yields). Black solid line shows the overall fit using sum of three Gaussian function including contribution from both symmetric and asymmetric fission. Green line is the mass distribution calculated using GEF code [17].